\documentclass[iop]{emulateapj}

\usepackage{graphicx}
\usepackage{grffile}
\graphicspath{{figures/}}

\newcommand{\phx}{\texttt {PHOENIX}}
\newcommand{\snia}{SN~Ia}
\newcommand{\sneia}{SNe~Ia}
\newcommand{\Mgt}{Mg\,{\sc ii}}
\newcommand{\Sit}{Si\,{\sc ii}}
\newcommand{\SitAbs}{Si\,{\sc ii}\ 6150\AA}
\newcommand{\SitLab}{Si\,{\sc ii}\ 6355\AA}
\newcommand{\St}{S\,{\sc ii}}
\newcommand{\Cat}{Ca\,{\sc ii}}
\newcommand{\Fet}{Fe\,{\sc ii}}

\begin{document}

\title{The Nature of the \SitAbs, \Cat\ HK, \Cat\ IR-triplet, and other Spectral Features in Supernova Type Ia Spectra}
\author{Daniel R. van Rossum}
\affil{Department of Astronomy and Astrophysics, University of Chicago, Chicago,IL 60637, USA}
\affil{Center for Computational Science, University of Chicago, Chicago,IL 60637, USA}

\keywords{stars: Supernovae: general - radiative transfer - methods: numerical}

\begin{abstract}
Spectra of Type Ia Supernovae (\snia) show both continuum-like and absorption-like line features.
The pseudo equivalent width (pEW) and Doppler shift of absorption-line-like features, such as \SitAbs, \Cat\ HK 3750\AA, and the \Cat\ IR-triplet 8150\AA\, quantities that are often associated with the optical depth and the velocity of a shell of interacting material, are important tools for interpreting and classifying the \snia\ spectra.
In this paper, we examine the nature of spectral features in \snia\ spectra using W7 model spectra and a technique that we call ``knock-out'' spectra.
We show that the P-Cyg profiles of \SitAbs\ and many other features are largely emission dominated, rather than absorption dominated, and that the concepts of ``absorption line'' and ``continuum'' are therefore not adequate to describe the nature of spectral features in \sneia\@.
Apparent absorption features, like \SitAbs, are frequently just coincidental troughs between two (or more) uncorrelated emission features.
In this situation, the pEW measured between these emission peaks is little related to the true strength of the presumed absorption feature.
Furthermore, using the same knock-out technique, we demonstrate how spectral features overlap each other at different times after the explosion.
This overlap distorts individual line profiles and affects measured absorption-line velocities.
With synthetic \snia\ ejecta stratifications that are tuned to match specific observations, the method presented in this paper can in principle be used to quantify the effects of specific line opacities on measured line velocity, line strength, and line blending, and improve the interpretation and informative value of observed \snia\ spectra.
\end{abstract}

\section{Introduction}
Supernovae of Type Ia (\snia) generally show very similar spectral and photometric behavior (e.g.\ \citealt{Filippenko97}).
Because of this interesting homogeneity (and the fact that these objects become extremely bright) \sneia\ have proven to be very important objects to study the universe at large distances \citep{Riess98, Perlmutter99}.
Yet despite the general similarity, there are an even more interesting photometric and spectroscopic diversities among \sneia.
Whereas photometrical diversity has long been studied \citep{Phillips93, Wang03}, good spectroscopic data have only recently become available in large numbers (CfA SN Program \citealt{Blondin12}; Berkeley \snia\ Program \citealt{Silverman12}).

One step towards understanding spectroscopic diversity is finding similarities and defining classes.
A number of spectral classification schemes for \snia\ have been proposed in literature.
These are based on measurements performed on conspicuous \Sit\ and \Cat\ absorption features present in all ``normal'' \citep{Branch93} \snia.
These absorption feature measurements include:
1) The feature strength, expressed as pseudo equivalent width (pEW, see \citealt{Folatelli04}) \citep{Branch06};
2) The line blue-shift, expressed as expansion velocity \citep{Wang09}; and
3) The rate of change in blue-shift, expressed as velocity gradient \citep{Benetti05}.

It is well known that atomic transition lines formed in a rapidly expanding atmosphere like \snia\ have P-Cyg profiles \citep{Lamers99}.
P-Cyg profiles generally show a superposition of an emission wing that is more or less symmetric around the transition's rest wavelength plus an absorption wing that is blue shifted.
But the precise shape of P-Cyg profiles in relativistically expanding atmospheres, and specifically the ratio of emission to absorption, strongly depends on details of the atmosphere \citep{Hutsemekers93}, including:
1) The radial extension of the line forming region;
2) The velocity gradient across the line forming region; and
3) The wavelength dependence of the ``continuum'' and its change across the line forming region.

Another fact that complicates the interpretation of \snia\ spectra is the large number of transition lines that leave their signature in the spectra in an indistinguishable way (i.e., smeared out by the large expansion velocities of \snia\ ejecta).
As shown by \cite{Kasen06b} and \cite{Kromer09} tens of millions of transitions must be included to calculate the \snia\ opacity accurately.
Even though some of these lines are stronger than others, the background opacity formed by the millions of weaker lines is not gray \citep{Kasen06b}, which leaves its signature on the profiles of those stronger lines.

Clearly, the line profiles in \snia\ spectra contain a large amount of information about the detailed atmospheric conditions (like the distribution of nuclear products), and understanding them better will increase the power of spectroscopic observations to improve our understanding of \snia\.
In this paper we will demonstrate that the P-Cyg profiles of \SitAbs\ and many other lines can be emission- instead of absorption-dominated well before peak brightness.
We also show the implications this has for interpreting the pEW and the Doppler shift of absorption-like line features.

\section{Methods}
As described in the previous section, a realistic reproduction of P-Cyg profiles requires a model  that describes the run of physical conditions over the geometrical extent of the line forming region.
In this paper we do not aspire to attain perfect reproductions of observed P-Cyg profiles but rather focus on understanding the general nature of spectral features as they are observed for \sneia.
For this purpose we calculate ``knock-out'' spectra from models in which a relatively small subset of atomic transition lines are removed from the opacity, but all other quantities (e.g.\ temperature, material density, electron density, atomic level population numbers) are kept fixed.
We then compare the knock-out spectra with the original spectra, which were calculated using the full, unaltered opacity.
The direct comparison between altered and unaltered opacity spectra shows what the signature of the removed subset of atomic transition line(s) on the full spectrum was.

The ``knock-out  technique is different from other, similar procedures that have been done or could be done.
For example, \cite{Bongard08} calculated ``single-ion'' spectra with the purpose of identifying the ions that shape the two so-called \Sit\ features in the 5000--7000\AA\ wavelength region.
Their method is similar in that they also redo a simulation with modified opacities and other quantities fixed.
The main difference is that, instead of perturbing the full opacities as little as possible, the single-ion opacities are qualitatively different (much lower) and therefore the single-ion spectra are not similar to the normal spectrum.

\cite{Kromer09} present an approach for monte carlo (MC) transport in which MC packets record the ion involved in the last line-emission interaction before the leaving the computational domain.
This procedure has the advantage that the additional information gained is obtained directly from a fully self-consistent simulation.
A limitation of this approach compared to the knock-out approach is that it provides no information about line absorption.

Another approach would be to alter the abundance of individual chemical elements, recalculate self-consistent light curves and spectra, and examine how spectral features change due to the change in abundance.
The temperature, electron density and population numbers will all change as a result of the changes in the opacity and the radiation field.
The changes in the properties of spectral lines can no longer be attributed to line opacities only but are the result of the complex interplay of line- and continuum-opacities, temperatures, electron densities, and ionization/excitation balance (through the electron density and temperature).
This approach really addresses the question of how the spectral lines would change if the composition of the ejecta changed, which is an important but very different question than the one we address here.

Each of these approaches has an important and valid use, but we believe that the knock-out approach provides the best insight into the question that we address in this paper: \emph{what is the signature of a particular atomic transition (or a set of transitions) on the full \snia\ spectrum.}

The subtle differences between knock-out spectra relative to the normal flux level require that model spectra be calculated with high precision and low noise levels.
In this paper, we use \phx, a state-of-the-art, general purpose stellar atmosphere code \citep{Hauschildt92, Hauschildt99, Hauschildt04, Hauschildt06, Vanrossum12}.
This code solves the special relativistic radiation transport problem with high accuracy using characteristic rays and operator splitting methods \citep{Olson87}.
The time evolution of synthetic \snia\ spectra is calculated self-consistently with \phx\ using the radiation energy balance (REB) method \citep{Vanrossum12}.
\phx\ with the REB method can in principle calculate synthetic \snia\ spectra in NLTE; however, NLTE is particularly challenging in the \snia\ problem, given the huge number of lines that play a role, and thus doing this is still a work in progress.
Therefore we use the LTE approximation in this paper.

Determining the density profile and the distribution of chemical elements in \snia\ ejecta from explosion models has been a highly active field of research for decades.
Since the first phenomenological 1-dimensional (1D) explosion model that successfully reproduced the basic observational characteristics of \snia\, W7 \citep{Nomoto84}, explosion models and simulations have incorporated more physics and use better numerical techniques (e.g.~\cite{Khokhlov91, Jordan08, Kasen09, Seitenzahl13, Kromer13, Long14}).
Yet the properties of the ejecta produced by these simulations often agree no better with those inferred from observations.
As a result, W7 therefore continues to be an important 1D reference model (e.g.~\cite{Kasen06, Kromer09, Vanrossum12, Wollaeger14}) that is useful for studying the nature of spectral features.  
The properties of the ejecta in the W7 model were not tuned to match a specific \snia\@.
For this reason, we compare our results to the \cite{Hsiao07} (hereafter Hsiao07) spectral templates rather than to the spectra of individual \snia.

\section{Results}
\begin{figure}
\centerline{\includegraphics[width=.49\textwidth]{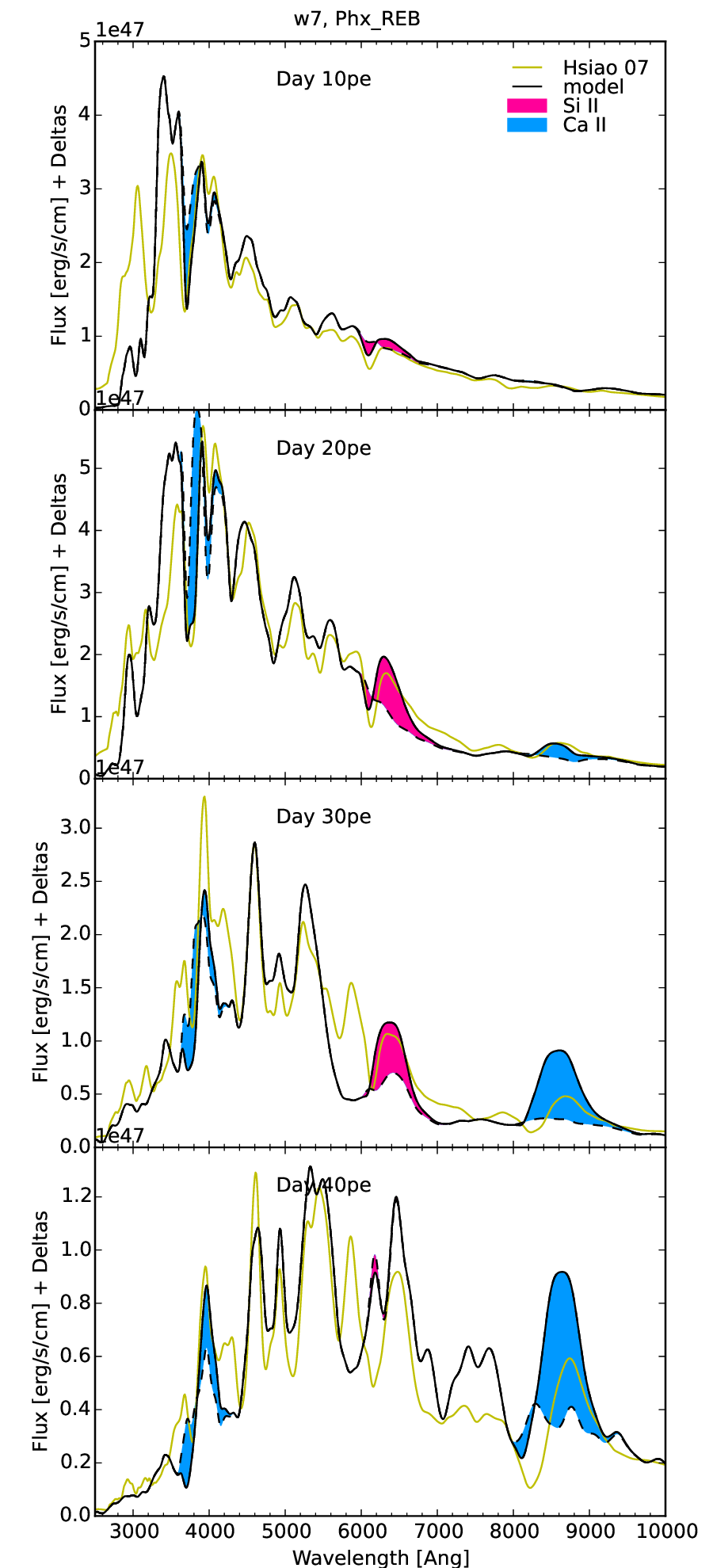}}
\caption{W7 model spectra (black solid line) compared to the Hsiao07 SN Ia spectral templates (yellow line) at different times post explosion (pe).
 The Hsiao07 templates are ``peak magnitude''-centered at day 18pe.
 The comparison uses only one free parameter to simultaneously match the template fluxes to the absolute fluxes from the model.
 The colored areas show the difference between the original model spectrum and the ``knock-out'' spectra (dashed black lines) for the \Sit\ 6355\AA, \Cat\ HK, and the \Cat\ IR triplet features.
 The colored areas above (below) the model spectrum show the absorption (emission) contribution to each of the features.
 The absorption part of the line profiles are generally much smaller than a pEW evaluation would suggest, and the emission strongly dominates the absorption wing in the \Sit\ 6355\AA\ and \Cat\ IRT features.
 This suggests that, when interpreting SN Ia spectra, the flux peaks should rather be identified with lines instead of the troughs.
 }\label{fig:lign}
\end{figure}

Figure~\ref{fig:lign} shows the W7 model spectra compared to the Hsiao07 templates at four different times post explosion (pe).
The peak template luminosity is aligned with the peak model luminosity at day 18pe.
A single scaling factor is applied to the unit-less template flux (simultaneously for all times) to globally match the model flux.
Apart from these two tightly constrained parameters, there is no freedom in the comparison.

In Figure~\ref{fig:lign} three important \snia\ features are examined with knock-out spectra: \Sit\ 6150\AA, \Cat\ HK 3750\AA, and the \Cat\ IR-triplet 8150\AA\@.
The lab wavelengths of the corresponding lines are \Sit\ 6347 and 6371\AA\ (commonly together referred to as 6355\AA), \Cat\ 3934 and 3969\AA, and \Cat\ 8498, 8542, and 8662\AA\@.
The colored areas below the regular spectrum represent features that are in emission in the normal (full) spectrum, while those above the regular spectrum represent features that are in absorption in the normal (full) spectrum.

\begin{figure*}
\centerline{%
\includegraphics[width=.53\textwidth]{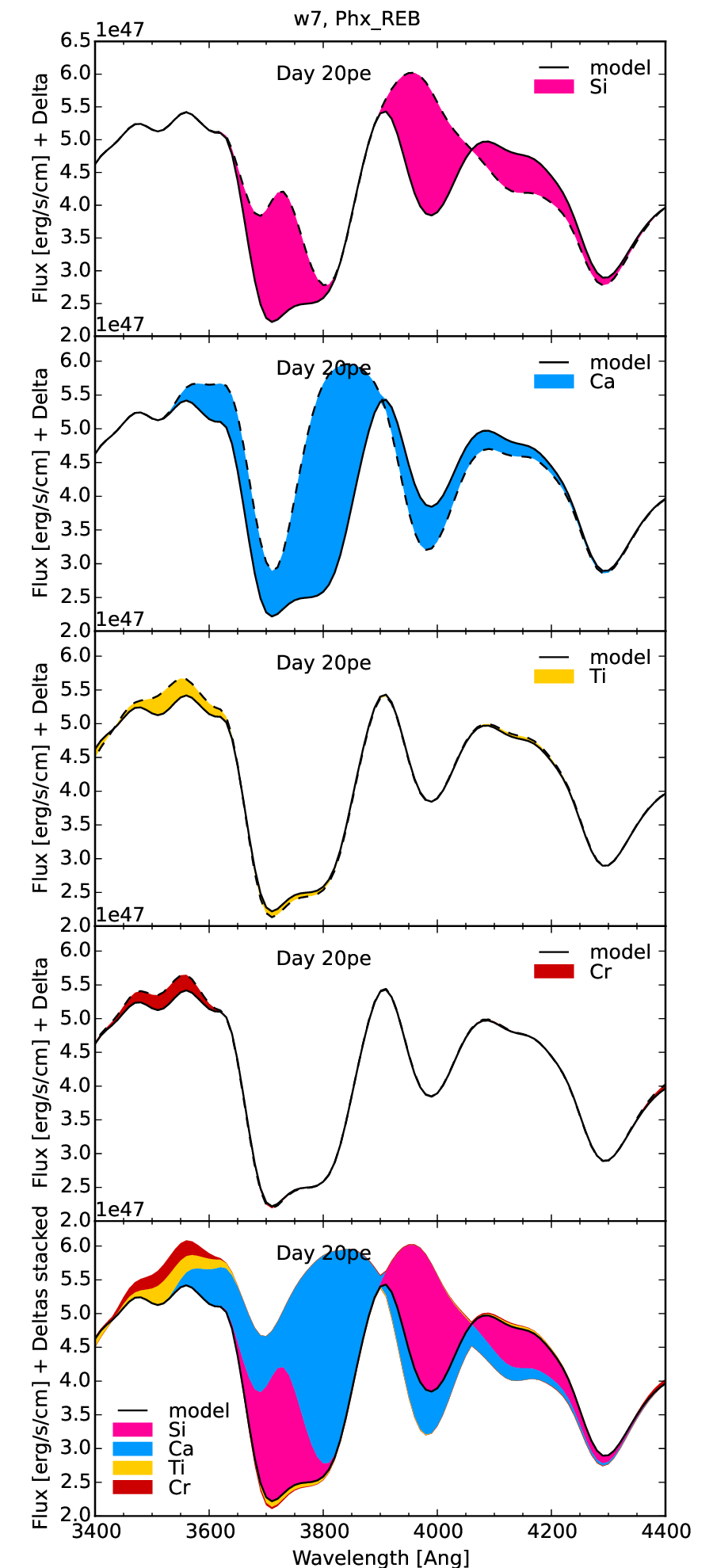}\hspace{-.3cm}
\includegraphics[width=.53\textwidth]{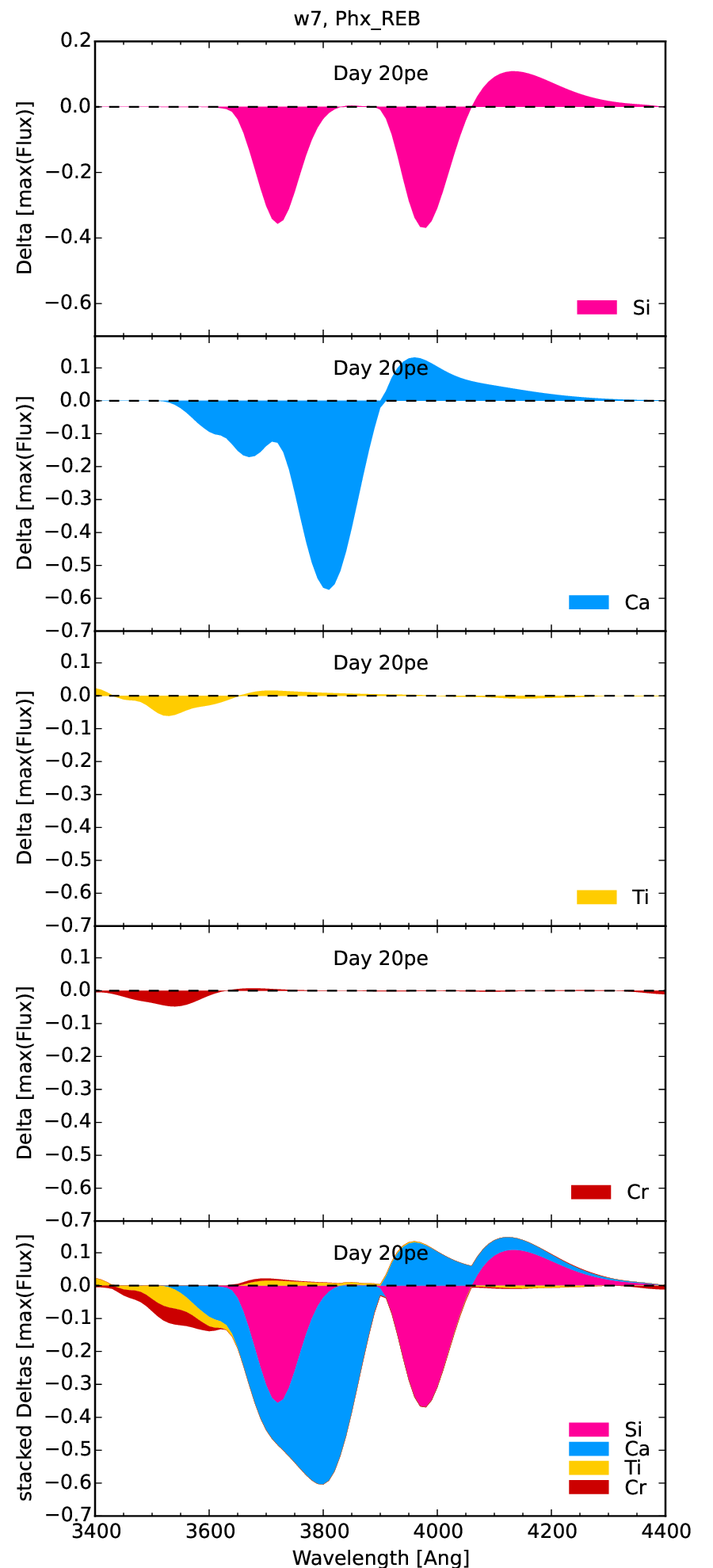}}
\caption{Combining the results from multiple knock-out spectra (four upper rows) into one plot (bottom row) helps condense the information and visualize which species play a role in shaping spectral features.
 The \emph{left column} shows the regular model spectrum (solid black line) at day 20 post explosion (close to peak brightness), zoomed in on the \Cat\ HK line.
 The four upper panels each show knock-out spectra for one chemical element at a time (dashed black lines), and the difference with the full spectrum is colored in.
 In the bottom panel the differences of each of the knock-out spectra are plotted on top of each other.
 In the \emph{right column} the differences are plotted relative to the maximum spectral flux level at that epoch, and are stacked in the same way in the bottom panel.
 This shows how in some wavelength regions the line opacities predominantly contribute to the full spectrum in absorption (from 3400--3900\AA; while in other regions, some contribute in absorption and some in emission (3900--4050\AA); and in still other regions, all of them contribute in emission (4050--4300\AA).
 We caution that the stacking of differences has no physical meaning, but is only done for visualization purposes.
 }\label{fig:0lin_stepbystep}
\end{figure*}

Figure~\ref{fig:0lin_stepbystep} shows how results from multiple knock-out spectra can be combined into a single plot.
We show the colored areas representing the knock-out spectra of lines or groups of lines on top of each other for two reasons.
First, this reduces the number of plots necessary to visualize the data.
Second, the stacked colored areas show how the shapes of different spectral features are formed by different line opacities or sets of line opacities.
It is important to note that there is no physical interpretation of the imaginary flux level at the top or the bottom of the stacked colored areas because spectra are not linear combinations of line profiles.

\begin{figure*}
\centerline{
\begin{minipage}[t]{.49\textwidth}
\centerline{\includegraphics[width=1.05\textwidth]{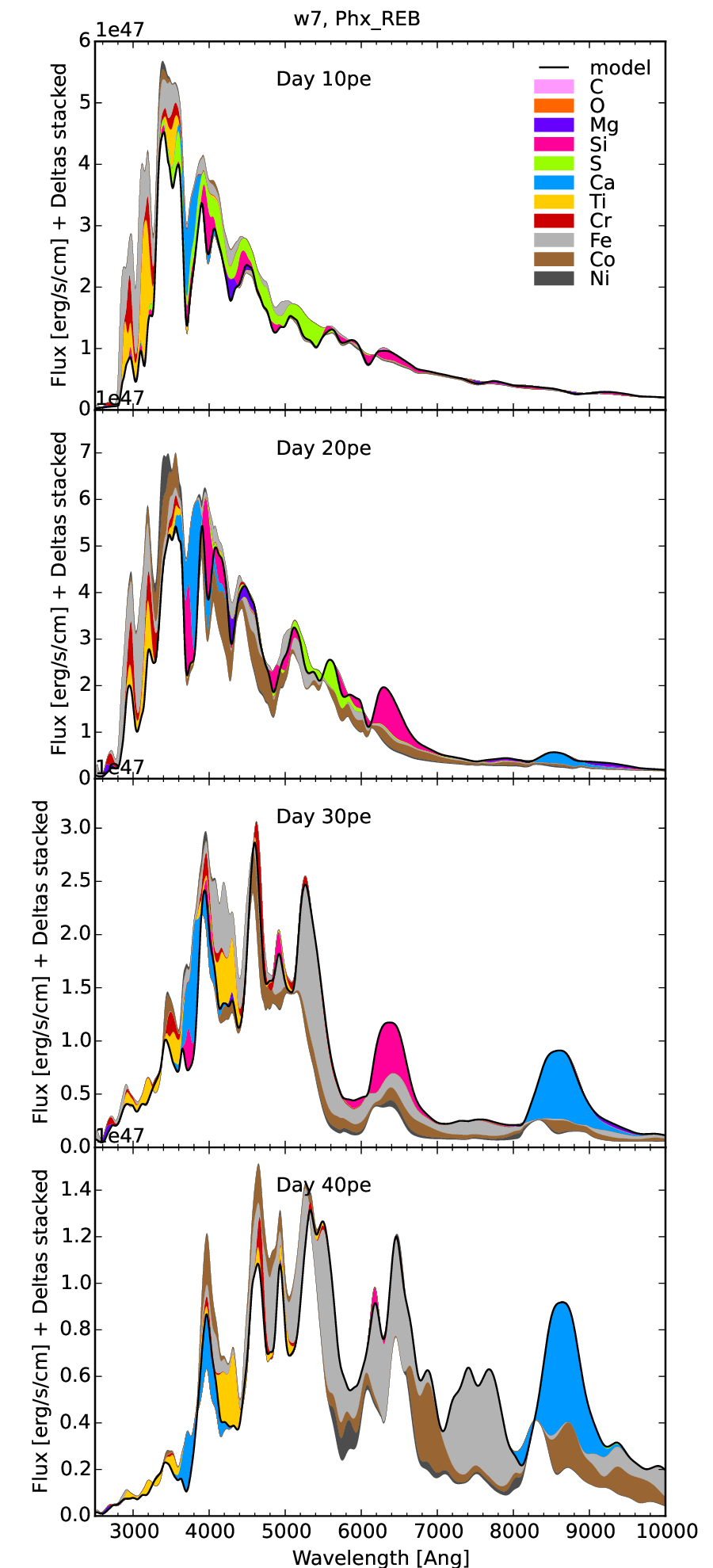}}
\caption{W7 model spectra (solid line) with knock-out spectra shown for different chemical species at 10, 20, 30, and 40 days post explosion (pe).
 Apparently, the spectra are largely \emph{emission dominated} longwards of 5500\AA\ and the pseudo-continuum level is significantly lower than the emission peaks on both sides of the absorption features\@.
 Ideally, pEW evaluations should take this into account.
 The important \SitLab\ feature is poorly reproduced by the W7 model (and by other models, see text), especially from day 30pe on.
}\label{fig:0lin_all}
\end{minipage}
\hspace{.2cm}
\begin{minipage}[t]{.49\textwidth}
\centerline{\includegraphics[width=1.05\textwidth]{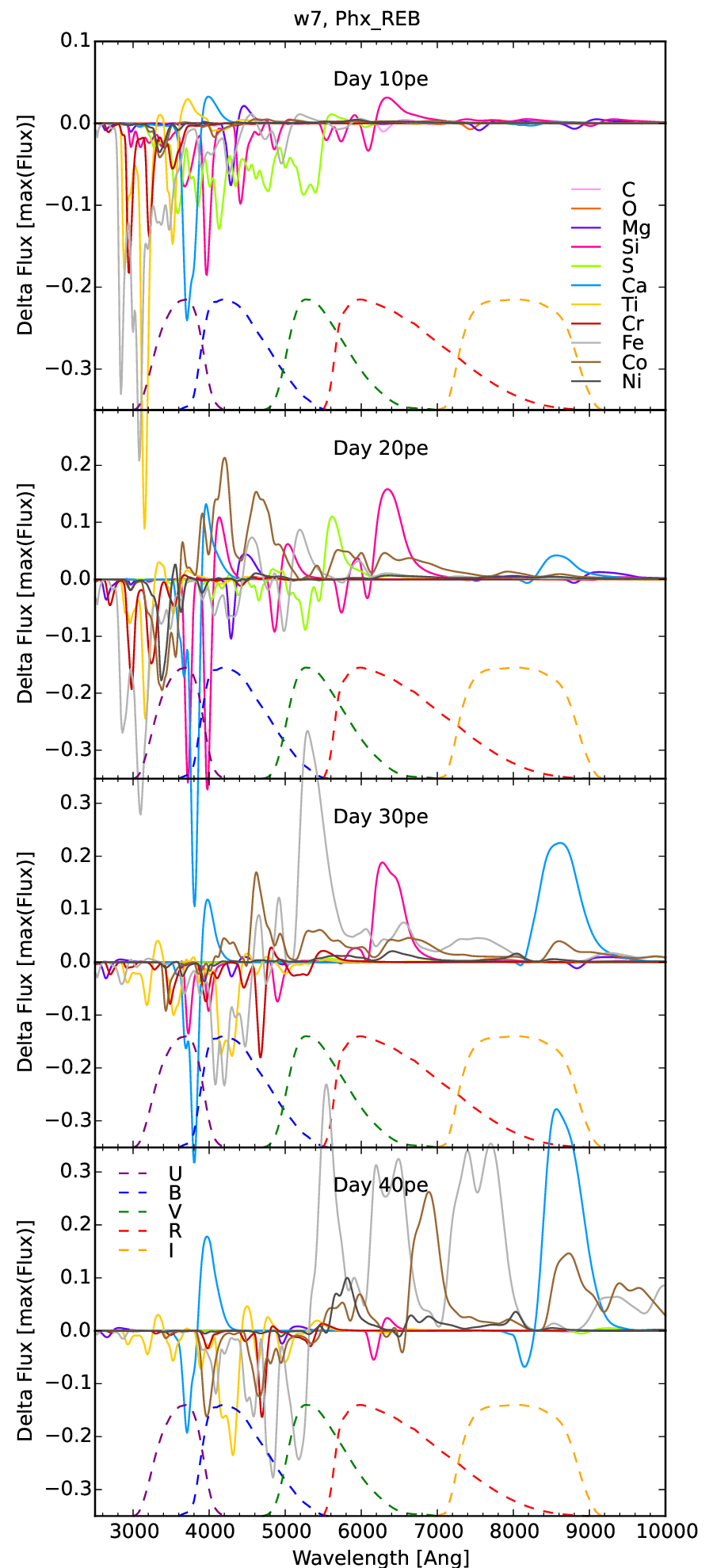}}
\caption{The same knock-out spectra for different chemical elements as shown in Figure~\ref{fig:0lin_all}, but plotted relative to the peak spectral flux level (see also Figure~\ref{fig:0lin_stepbystep}) without stacking.
 The dashed curves show the UBVRI \citealt{Bessell90} band filter functions.
 The important \Sit\ 6150\AA\ and \Cat\ HK features, and many others, show emission wings that are often stronger than the respective absorption wings\@.
 The \Cat\ HK feature (like many others) is not an isolated feature, but is instead blended with contributions from Si, Co, Ti, and S.
 This compromises measurements of both the feature strength and velocity.
 In all four epochs shown here, the U-band and the B-band contain contributions from many different elements, more so than others.
}\label{fig:0lin_diffeach}
\end{minipage}
}
\end{figure*}

Figure~\ref{fig:0lin_all} shows the same W7 model spectra at day 10, 20, 30, and 40 post explosion as in Figure~\ref{fig:lign}, but instead of showing three specific features, knock-out spectra are plotted for all of the atomic line opacities that contribute significantly to the shape of the spectrum, with different colors corresponding to different chemical elements.
In order to stress that the stacked colored areas in Figure~\ref{fig:0lin_all} have no physical meaning, Figure~\ref{fig:0lin_diffeach} shows the same knock-out spectra plotted relative to the peak spectral flux level and without stacking.
For reference, the figure also shows the \cite{Bessell90} UBVRI filter functions, arbitrarily normalized.

\begin{figure}
\centerline{\includegraphics[width=.51\textwidth]{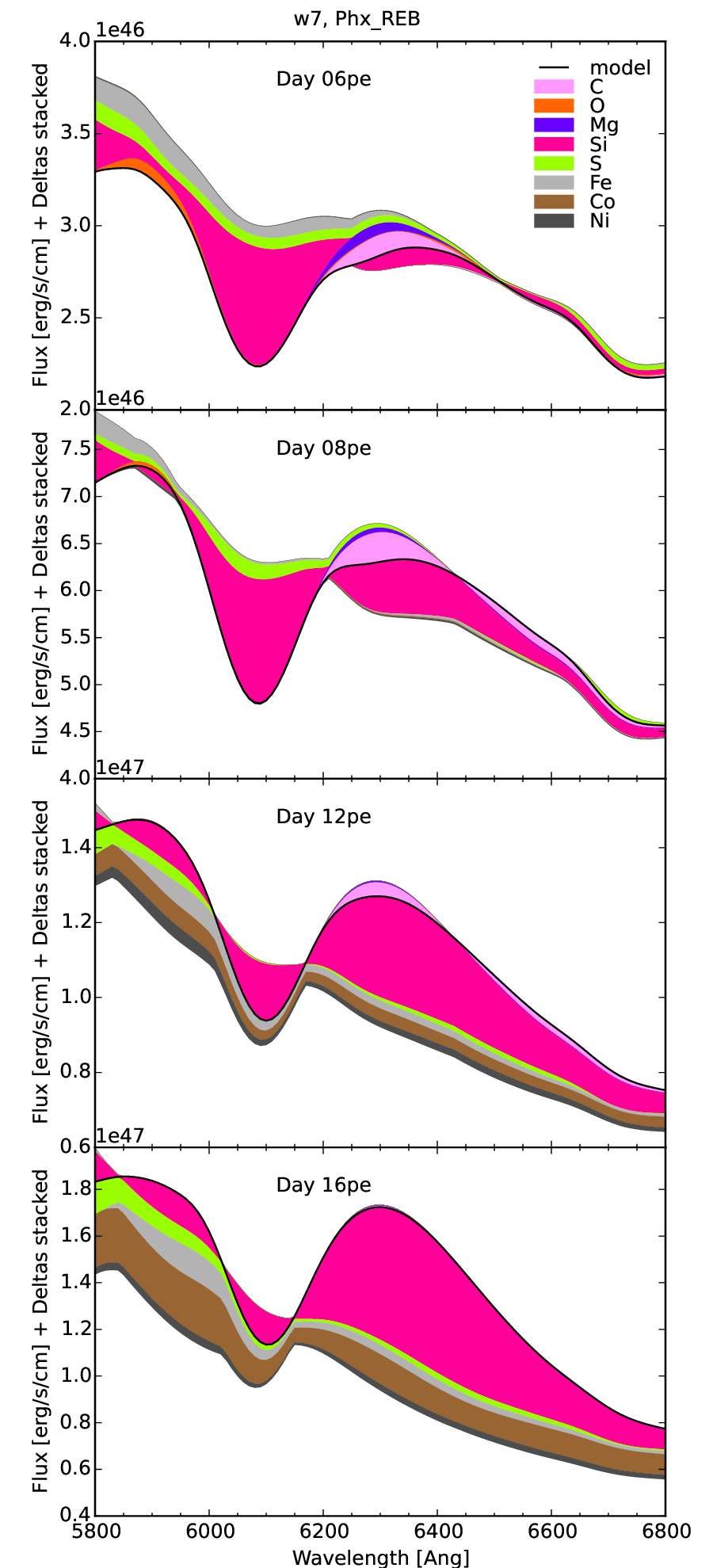}}
\caption{Zoomed-in spectra showing the \SitAbs\ feature of Figure~\ref{fig:0lin_all}, but for different times: days 6, 8, 12, and 16 post explosion.
 The true absorption signature of this important \Sit\ feature is smaller than the total depth of the trough, and decreases over time. 
Before peak brightness, the emission wing on the blue side of the 6150\AA\ trough is mainly affected by O, Si, S, Fe, and Co line opacities, while that on the red side is mainly affected by C, Mg, Si, and Co line opacities.
 The wavelength at which the trough minimum appears depends not only on the velocity of the \SitLab\ line, but also on how other features are blended in (including emission from the \Sit\ 5972\AA\ line).
 }\label{fig:0lin_si2_all}
\end{figure}

Figure~\ref{fig:0lin_si2_all} zooms in on the \SitAbs\ feature at 6, 8, 12, and 16 days post explosion. 
In contrast to Figure~\ref{fig:lign}, which shows the knock-out spectrum for the \SitLab\ line opacity only, here the combined effect of all of the Si line opacities is shown in pink, similar to Figure~\ref{fig:0lin_all}. 
The \SitAbs\ feature is one of the most distinct features in \snia\ spectra, and is often used to make line velocity and pEW measurements. 
It can be seen that the shape of this feature is determined not only by the \SitAbs\ line opacity itself, but also by C, O, Mg, S, Fe, and Co line opacities and by other Si line opacities like \Sit\ 5972\AA\@. 
These blends of other lines affect both the velocity at which the profile minimum occurs and the pEW of the line. 
In addition, over time the profile of the \SitAbs\ line itself develops an emission wing that is stronger than its absorption component. 
Thus the true strength of the \SitAbs\ absorption feature is smaller than the pEW of the whole trough, which is partially formed by multiple, correlated, small, emission-line wings on either side.

\begin{figure}
\centerline{\includegraphics[width=.5\textwidth]{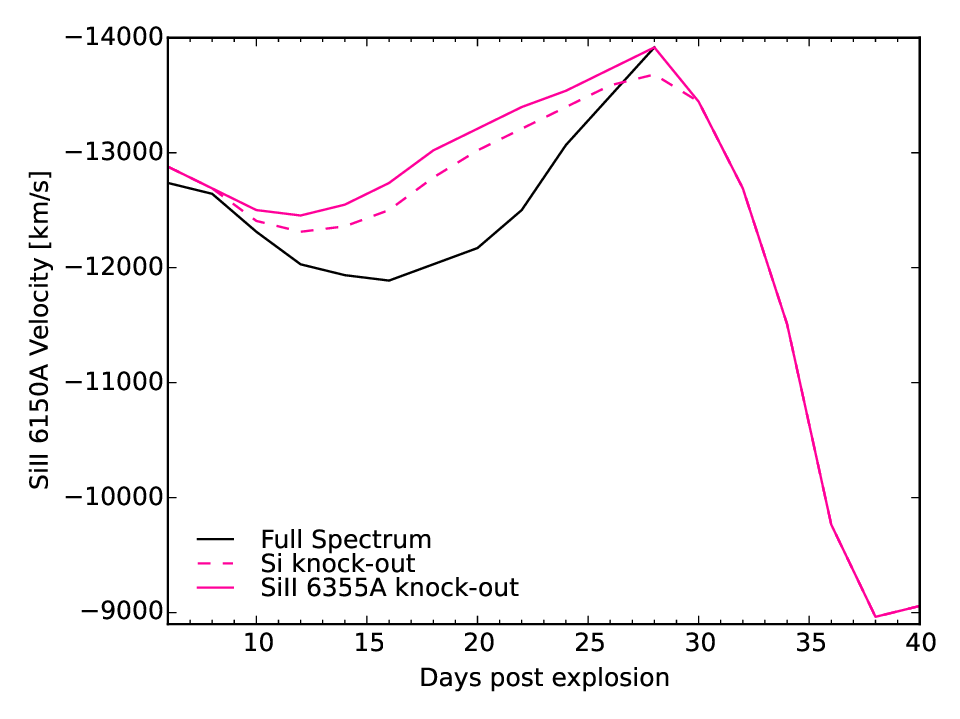}}
\caption{The absorption velocity of the \SitAbs\ feature measured in three ways:
 1) in the regular spectrum (black curve),
 2) in the Si knock-out curve of Figure~\ref{fig:0lin_diffeach} (dashed pink curve), and
 3) in a \SitLab\ only knock-out curve (solid pink curve).
 The solid pink curve is a clean measure of Si absorption velocity, measured for an isolated single feature.
 The dashed pink curve shows that the emission wing of the neighboring \Sit\ 5972\AA\ line reduces the apparent absorption velocity, although the effect is small.
 After day 28pe, the \Sit\ trough disappears in the regular spectra and can no longer be measured in the W7 model spectra.
}\label{fig:si2_v0}
\end{figure}

The effects of line blending on the absorption velocity of the \SitAbs\ trough are demonstrated in Figure~\ref{fig:si2_v0}.
The feature velocity as measured in the regular spectrum is lower than the actual line velocity of the isolated \SitLab\ line feature.
>From Figure~\ref{fig:0lin_diffeach} and Figure~\ref{fig:0lin_si2_all} it can be inferred that line emission from S, Fe, and Co lines lift up the left side of the \SitLab\ profile more than the right side.
>From day 28pe on the \SitLab\ feature can no longer clearly distinguished from neighboring features in the W7 model spectra.

\begin{figure}
\centerline{\includegraphics[width=.5\textwidth]{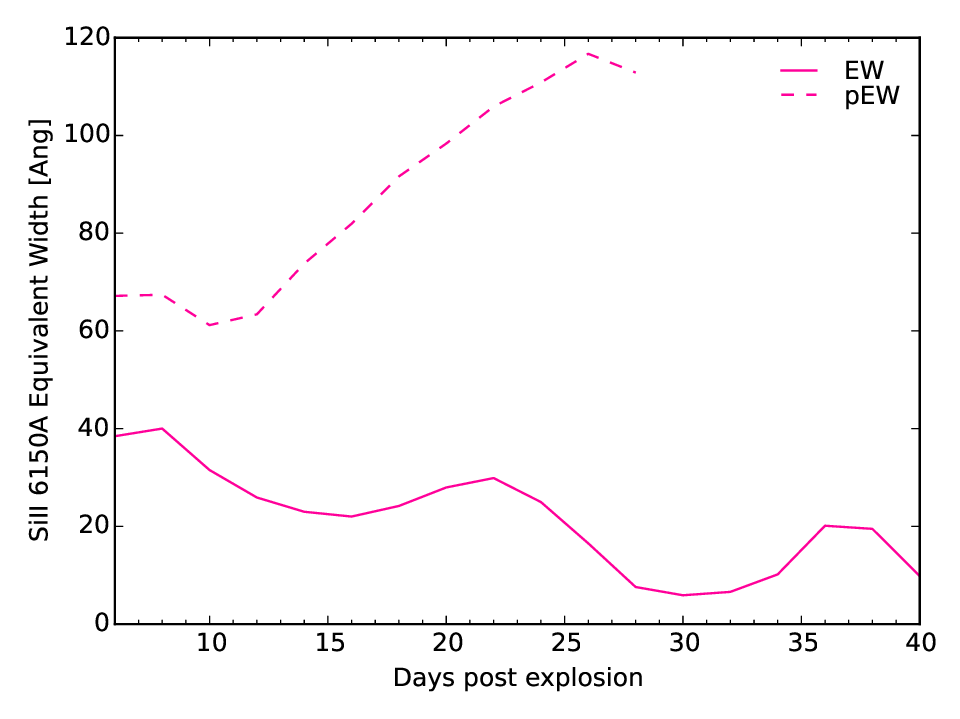}}
\caption{The temporal evolution of the pseudo Equivalent Width (pEW) and the true Equivalent Width (EW) of the \SitAbs\ absorption feature.
 The pEW is evaluated using a linear pseudo continuum that runs between the flux peaks on both sides.
 The true EW is measured using the true continuum level that is determined with knock-out spectra.
 The true EW is significantly smaller than the pEW but, more importantly, the trend is different between the two.
 From day 28pe on the \SitLab\ feature can no longer clearly distinguished from neighboring features in the W7 model spectra.
}\label{fig:si2_pew}
\end{figure}

Figure~\ref{fig:si2_pew} shows two different measurements of the strength of the \SitAbs\ feature.
The pEW measurement is agnostic about the nature of the feature and assumes a linear run of the continuum between the flux peaks on either side of the absorption feature.
The true equivalent width (EW) measurement evaluates the true absorption of the \SitLab\ line only, isolated from other, unrelated features, and using the knowledge of the true continuum for this particular line, as it can be determined using knock-out spectra for this line.
The pEW values are much larger than the EW values but, more importantly, the trend is different between the two.

\begin{figure}
\centerline{\includegraphics[width=.5\textwidth]{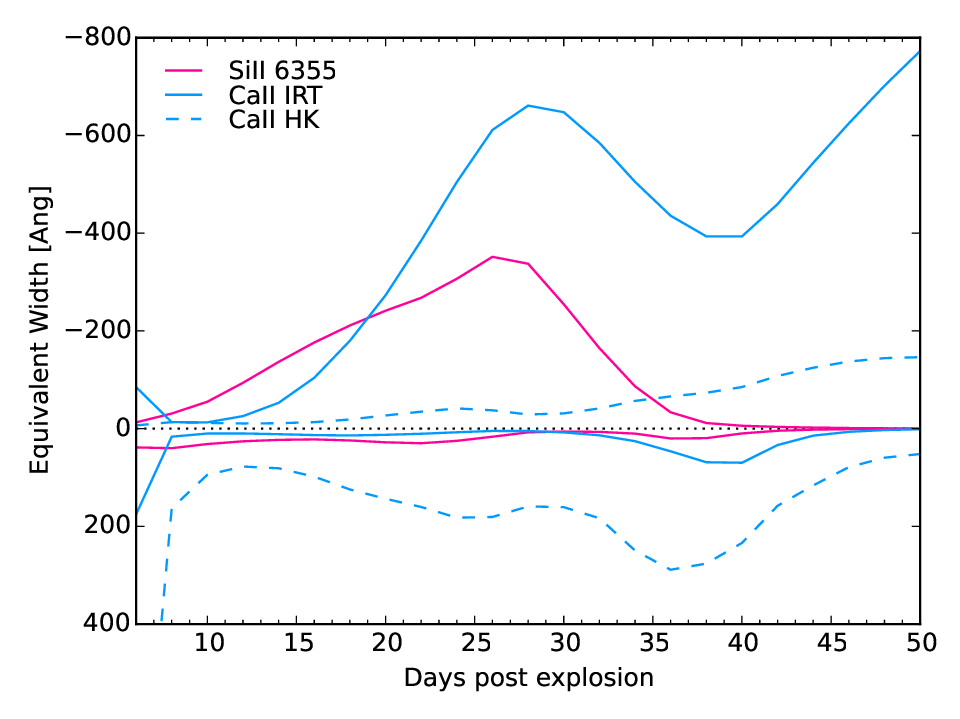}}
\caption{The evolution of the true equivalent width (EW) of three spectral features over time.
 The EW is computed using the knock-out spectra from individual lines, i.e.\ the line profiles are evaluated versus the true run of the continuum flux level over each line profile.
 Each of the three features has a positive and a negative curve corresponding to the EW measures of the absorption and emission components, respectively.
 The \SitLab\ and the \Cat\ IRT line profiles are emission dominated from approximately day 10pe on.
 The \Cat\ HK line profile is absorption dominated before day 44pe.
}\label{fig:si2ca2_ew}
\end{figure}

Figure~\ref{fig:si2ca2_ew} presents the evolution of the EW measure of the \SitLab, \Cat\ IRT, and \Cat\ HK features.
Similar to Figure~\ref{fig:si2_pew}, the EW is calculated using the run of the known true continuum level over the line profiles, as determined from the knock-out spectra for each of these lines.
In the W7 model spectra, the \SitLab\ and the \Cat\ IRT line profiles are emission dominated from approximately day 10pe on.
The \Cat\ HK line profile is absorption dominated before day 44pe.

\subsection{Absorption features}
\paragraph{\SitAbs\ absorption}
On day 10pe, the absorption trough around 6150\AA\ is dominated by \Sit\.
On day 20pe, the \Sit\ feature has mostly changed to emission and is significantly supported by Co emission.
This trend continues on day 30pe, before Fe emission (at nearly the same wavelength) completely takes over at day 40pe.                                                                              
\paragraph{\St\ ``W'' absorption}
On day 10pe, the absorption trough around 5700\AA\ is clearly dominated by \St\@.
On day 20pe, it is heavily blended with Co and Fe emission, and on day 30pe \St\ has vanished.

\paragraph{\Mgt\ absorption}
On day 10pe, the \Mgt\ absorption around 4300\AA\ is blended with Si, S, and Fe absorption.
On day 20pe, Fe absorption and Co emission dominate \Mgt, and on day 30pe \Mgt\ has vanished.

\paragraph{\Cat\ absorption}
The \Cat\ HK absorption around 3750\AA\ at day 10pe has contributions from Si and S absorption and Ti emission.
On day 20pe and 30pe, the \Cat\ HK trough is significantly deepened by Si absorption, obviously affecting its pEW (note that this problem does not affect the true EW values plotted in Figure~\ref{fig:si2ca2_ew}).
This also changes the absorption velocity.

The \Cat\ IRT feature appears mainly in emission.

\paragraph{\Fet\ 5000\AA\ blend}
The wide absorption trough around 5000\AA\ is commonly identified as the ``\Fet\ blend'' (e.g.\ \cite{Folatelli04, Hachinger06}).
Figure~\ref{fig:0lin_all} and~\ref{fig:0lin_diffeach} show, however, that the Fe is weaker than contributions from other elements at day 10, 20, and 30pe.
Only as late as day 40pe, this feature is really dominated by Fe absorption.

\section{Discussion and Conclusions}
\subsection{Absence of an important emission peak}
Generally, W7 provides a reasonably good fit to the Hsiao07 templates.
There is, however, one important discrepancy.
The model does not produce the emission peak around 5800\AA\ that is observed on day 20, 30, and 40pe.
This discrepancy is also seen in the  spectra for W7 calculated using other radiation transport codes (e.g.\ \citealt{Kasen06, Kromer09, Wollaeger14}).
This is problematic because this emission peak is responsible for the blue side of the \SitAbs\ trough important for line velocity and pEW measurements.

What makes this problem even more interesting is that other \snia\ explosion models do not produce this emission peak either (e.g.\ \citealt[Figure 2b]{Kasen09}; \citealt[Figure 3]{Roepke12}; \citealt[Figure7]{Sim13}).
There are two possible reasons why this emission peak is absent from the models: 1.\ the chemical element(s) responsible for the observed peak is(are) under-represented in the explosion models, 2.\ the opacity data for the elements present in the models are missing one or more important transition lines\footnotemark{}.
Note that NLTE effects do affect the shape of \snia\ spectra \citep{Baron96}, but are unlikely to explain discrepancies that are as large as this missing emission peak.

Since the origin of the observed emission line is unknown and it shapes the blue side of the \SitAbs\ absorption trough, special care must be taken in interpreting the pEW of this important trough.

\subsection{Blended features}
Line features are known to strongly blend in \snia\ spectra, but it is still common to speak of \emph{identified} features, named after the predominant contributor.
Figures~\ref{fig:0lin_all} and~\ref{fig:0lin_diffeach} show the degree of blending of the spectral lines in the model spectra as a function of time.
The degree of blending depends on the detailed atmospheric conditions, which vary between \sneia.
The W7 results presented here provide a qualitative picture of the blending effects.
When combined with synthetic \snia\ ejecta configurations (density and composition) that are tuned to fit the spectra of a particular \snia, the knock-out spectra approach presented in this paper would allow the degree of blending to be determined quantitatively,
increasing the information that can be gleaned from \snia\ spectra.

\subsection{The \SitAbs\ feature}
The \SitAbs\ feature is the strongest feature in many \snia\ spectra and is the feature that is most widely used to characterize and classify \snia\ spectra.  It therefore deserves special attention.
Figure~\ref{fig:si2_v0} compares the velocity of the \SitLab\ absorption line with that of the 6150\AA\ trough.
Thus the figure compares the velocity of isolated Si with the value measured using the full spectrum.
At early times, the \SitAbs\ trough is filled on the blue side by emission from the next bluer \Sit\ line, leading to lower measured velocities, but the effect is small, amounting to only a few percent.\\
Figure~\ref{fig:si2ca2_ew} shows that \Sit\ emission gradually declines from day 26pe on.
At the same time, the peak of the \Sit\ emission is seamlessly taken over by Fe\,{\sc ii} emission at a slightly larger wavelength.
This reduces the pEW, and reduces the measured \Sit\ velocity and increases its measured gradient, at late times compared to their actual values.

For these reasons, correlating the measured \SitAbs\ velocity at late times with values at around peak luminosity or earlier (see, e.g.\ \citealt{Howell06, Blondin12}) could be problematic, since the early- and late-time measured values of the velocity do not have the same physical origin.
The same problems may apply to the definition of the velocity gradient definition used in \cite{Benetti05}.
The improved velocity gradient definition used in \cite{Blondin12} limits the use of \SitAbs\ velocities to day 10 post-maximum brightness (approximately day 28pe for W7), and thus partly avoids these problems.

\subsection{The nature of spectral features in \snia\ spectra}
In this subsection we discuss why the concepts of absorption-line identifications and ``continuum'' can be misleading terms for interpreting \snia\ spectra.

\paragraph{Line Identifications}
In the interpretation of \snia\ spectra, line-identification labels are often associated with apparent absorption troughs, and the peaks in the flux between the troughs are used to define the ``continuum'' level.
However, many of the features in \snia\ spectra are due predominantly to emission.
Figure~\ref{fig:0lin_diffeach} shows that, for wavelengths larger than 5400\AA\, the spectral features generally are emission dominated (except possibly \Sit\ at very early times).  Even at day 20pe (close to peak luminosity), emission contributes significantly to the flux level down to wavelengths as short as 3800\AA\@.
The widths and depths of these absorption troughs depend strongly on the strengths and separation of the adjacent emission features.

\paragraph{The Continuum Flux Level}
Estimating a realistic pseudo continuum for a \snia\ spectrum is not an easy task.
What can be said is that estimating the pseudo continuum using a line connecting adjacent observed flux peaks in the spectrum is unrealistic for many features at most evolution times, since line emission is strong in \snia\ spectra.

\paragraph{Definition of \snia\ feature strength}
\snia\ ejecta configurations (density and composition) that are tuned to fit the observed spectra of a specific \snia\ would, in principle, allow the pseudo continuum to be determined for any spectral feature, using the method presented in this paper.
However, such tuned ejecta configurations are not yet generally available.

In the meantime, it might be better to measure the strength of spectral features using the height difference of P-Cyg profiles, rather than the pEW method.
The strength of a spectral feature would then be defined as the difference between the flux at the minimum of the trough and the flux at the maximum of the peak on the red side of the trough, normalized to the mean of the two fluxes.
The emission peak on the blue side of the trough, whose strength is unrelated to the strength of the feature of interest, would thus be excluded, potentially giving a more robust and informative measure of feature strength.
Note that, using this definition, the strength of a spectral feature is a dimensionless quantity that does not use the width of the feature, which is loosely related to the absorption velocity.
Consequently, the measured strength and velocity of a velocity feature may be more independent.
Applying this new measure of the strength of spectral features to large \snia\ spectral datasets can show whether its use strengthens the correlation between it other spectroscopic or photometric properties, or not, and how its use affects classification schemes based on line strengths.

In future work, we will extend the method presented in this paper by using synthetic \snia\ ejecta configurations (density and composition) tuned to produce synthetic spectra that match the spectra of specific \sneia.
Such an approach, using parameterized radiation transfer calculations, based on both W7 and 2D and 3D simulations of more recent explosion models, will help make it possible to infer information from observed \snia\ spectra in a more physically self-consistent way.

\hspace{.5cm}
\acknowledgements{}
We thank Don Lamb, Dean Townsley, and Sean Couch for helpful suggestions and comments on the manuscript of this paper.
This work is supported in part by the National Science Foundation under grant AST-0909132, and under grant PHY-0822648 for the Physics Frontier Center ``Joint Institute for Nuclear Astrophysics'' (JINA).

\bibliography{bibAstro,bibSNexp,bibSNobs,bibRT}

\end{document}